\PassOptionsToPackage{unicode}{hyperref}
\PassOptionsToPackage{hyphens}{url}
\documentclass[
]{article}
\usepackage{lmodern}
\usepackage{amssymb,amsmath}
\usepackage{ifxetex,ifluatex}
\ifnum 0\ifxetex 1\fi\ifluatex 1\fi=0 
  \usepackage[T1]{fontenc}
  \usepackage[utf8]{inputenc}
  \usepackage{textcomp} 
\else 
  \usepackage{unicode-math}
  \defaultfontfeatures{Scale=MatchLowercase}
  \defaultfontfeatures[\rmfamily]{Ligatures=TeX,Scale=1}
\fi
\IfFileExists{upquote.sty}{\usepackage{upquote}}{}
\IfFileExists{microtype.sty}{
  \usepackage[]{microtype}
  \UseMicrotypeSet[protrusion]{basicmath} 
}{}
\makeatletter
\@ifundefined{KOMAClassName}{
  \IfFileExists{parskip.sty}{%
    \usepackage{parskip}
  }{
    \setlength{\parindent}{0pt}
    \setlength{\parskip}{6pt plus 2pt minus 1pt}}
}{
  \KOMAoptions{parskip=half}}
\makeatother
\usepackage{xcolor}
\IfFileExists{xurl.sty}{\usepackage{xurl}}{} 
\IfFileExists{bookmark.sty}{\usepackage{bookmark}}{\usepackage{hyperref}}
\hypersetup{
  pdftitle={A conditional independence test for causality in econometrics},
  pdfauthor={Jaime Sevilla; Alexandra Mayn},
  hidelinks,
  pdfcreator={LaTeX via pandoc}}
\usepackage[margin=1in]{geometry}
\usepackage{color}
\usepackage{fancyvrb}

\DefineVerbatimEnvironment{Highlighting}{Verbatim}{commandchars=\\\{\}}
\usepackage{framed}
\definecolor{shadecolor}{RGB}{248,248,248}
\newenvironment{Shaded}{\begin{snugshade}}{\end{snugshade}}

\newcommand{\CommentTok}[1]{\textcolor[rgb]{0.56,0.35,0.01}{\textit{#1}}}

\newcommand{\ControlFlowTok}[1]{\textcolor[rgb]{0.13,0.29,0.53}{\textbf{#1}}}
\newcommand{\DataTypeTok}[1]{\textcolor[rgb]{0.13,0.29,0.53}{#1}}
\newcommand{\DecValTok}[1]{\textcolor[rgb]{0.00,0.00,0.81}{#1}}

\newcommand{\FloatTok}[1]{\textcolor[rgb]{0.00,0.00,0.81}{#1}}

\newcommand{\KeywordTok}[1]{\textcolor[rgb]{0.13,0.29,0.53}{\textbf{#1}}}
\newcommand{\NormalTok}[1]{#1}
\newcommand{\OperatorTok}[1]{\textcolor[rgb]{0.81,0.36,0.00}{\textbf{#1}}}
\newcommand{\OtherTok}[1]{\textcolor[rgb]{0.56,0.35,0.01}{#1}}

\newcommand{\StringTok}[1]{\textcolor[rgb]{0.31,0.60,0.02}{#1}}

\usepackage{graphicx,grffile}
\makeatletter
\def\maxwidth{\ifdim\Gin@nat@width>\linewidth\linewidth\else\Gin@nat@width\fi}
\def\maxheight{\ifdim\Gin@nat@height>\textheight\textheight\else\Gin@nat@height\fi}
\makeatother
\setkeys{Gin}{width=\maxwidth,height=\maxheight,keepaspectratio}
\makeatletter
\def\fps@figure{htbp}
\makeatother
\providecommand{\tightlist}{%
  \setlength{\itemsep}{0pt}\setlength{\parskip}{0pt}}
\usepackage{booktabs}
\usepackage{float}

\title{A conditional independence test for causality in econometrics}
\author{Jaime Sevilla \and Alexandra Mayn}
\date{24 June, 2021}

\begin{document}
\maketitle

\textbf{ABSTRACT}

The Y-test is a useful tool for detecting missing confounders in the
context of a multivariate regression.However, it is rarely used in
practice since it requires identifying multiple conditionally
independent instruments, which is often impossible. We propose a
heuristic test which relaxes the independence requirement. We then show
how to apply this heuristic test on a price-demand and a firm
loan-productivity problem. We conclude that the test is informative when
the variables are linearly related with Gaussian additive noise, but it
can be misleading in other contexts. Still, we believe that the test can
be a useful concept for falsifying a proposed control set.

\begin{center}\rule{0.5\linewidth}{0.5pt}\end{center}

\textbf{INTRODUCTION}

The problem of attributing causal meaning to statistical correlations is
both well known and important. As a motivating example, we recall
Messerli (2012), where it is concluded that the chocolate consumption of
a country was a cause of the number of nobel prizes its scientists
earned - where presumably what happened is that both were related by an
unobserved confounder such as the country's wealth.

Traditionally, questions of causality in econometrics have been
addressed through structural equation modeling. The gold standard for
causal inference from observational data are natural experiments, where
researchers identify an instrument that affects variation of the
exposure and use it to measure the unconfounded causal effect of the
exposure on the outcome (Reiersöl 1945).

However, natural experiments require us to make some assumptions about
the causal structure of the system we are studying. In particular, they
require us to assume that the instrument does not affect the outcome we
are interested in, except indirectly through its influence on the
exposure (the so-called \emph{exclusion restriction}).

One way to address this problem is through overidentification tests,
where a causal measurement is performed using multiple different
instruments to verify that the results obtained are consistent (Sargan
1958). In this article we will present a complementary approach, also
employing multiple instruments to verify the validity of a natural
experiment.

More recently, graphical models were introduced as an complement to
structural equation modelling to represent and discover causality. See
for example Glymour, Zhang, and Spirtes (2019) for a contemporary review
of graphical causal discovery methods.

Despite the advances in the last two decades, many insights from the
field of graphical causal discovery have gone unnoticed in econometrics.
Here we discuss one such insight: y-structure tests (Mani, Spirtes, and
Cooper 2012).

When we have access to multiple conditionally independent instruments,
y-structure tests can be used to falsify a proposed controlling set.
Hence they can be used to infer causality without relying on assumptions
of exclusion restriction. We explain how and illustrate the method using
synthetic data in section 1.

In practice, the requirement of conditional independence of the
instruments often does not hold, so the Y-test in its classical form
cannot be applied. We address this limitation by proposing a heuristic
version of the Y-test which does not assume conditional independence of
the instruments in section 2, which we test on synthetic data.

We illustrate the application of the heuristic test using data from a
cigarette demand vs price elasticity study in section 3, and using data
from a bank loan vs firm productivity study in section 4.

Section 5 wraps up the article with a discussion of our results and
remaining open questions.

\hypertarget{y-structure-tests}{%
\subsection{1. Y-structure tests}\label{y-structure-tests}}

Let us now examine the problem Y-tests are used to solve.

We have a matrix of observational data, where each row is an iid sample
of the system and each column contains the observations corresponding to
a variable of the system.

The observed variables are entangled in a directional web of cause and
effect, so that interventions manipulating a variable result in changes
of the variables that are causally downstream from them.

In particular, we are interested in the relation between an exposure
variable \(X\) and an outcome variable \(Y\) - our goal is to determine
whether the outcome \(Y\) is causally downstream from the exposure
\(X\), and if so, what is the strength of the causal effect of the
exposure \(X\) on the outcome \(Y\). In other words, how much would a
marginal, exogenous increase of \(X\) affect \(Y\)?

In order to do so we aim to verify that a group of variables \(S\),
which we will call the \emph{context}, blocks all non causal paths
between \(X\to Y\). If we had such a context \(S\), we could measure the
causal effect of \(X\) on \(Y\) by regressing \(Y\) on \(\{X\} \cup S\),
and using the coefficient of partial correlation \(\rho_{XY\cdot S}\) as
our estimate.

We assume that the causal system we study is acyclic - there are no
causal loops - and we assume that the data is faithful to its underlying
graph, i.e.~conditional independences are an accurate reflection of the
causal structure we are studying. See appendix A for a more precise
statement of these conditions.

Importantly, we do not assume that all variables are observed - there
might be unobserved common causes. Y-tests will help us rule out this
possibility.

For simplicity we will assume a system with linear causal relationships
and gaussian additive noise. Because of this we will be able to measure
strength of (conditional) statistical dependence as (partial) linear
correlations. It is possible to apply Y-tests to non-linear,
non-gaussian causal systems, but we would need to use an alternate way
of measuring (conditional) statistical dependence, such as (conditional)
mutual information tests.

As an example, we will first work with some synthetic data. Let's
suppose, for example, that we are working with a system whose generative
process can be approximately modeled as follows:

\begin{Shaded}
\begin{Highlighting}[]
\NormalTok{n =}\StringTok{ }\DecValTok{100}
\NormalTok{A <-}\StringTok{ }\KeywordTok{rnorm}\NormalTok{(n)}
\NormalTok{Z <-}\StringTok{ }\NormalTok{A }\OperatorTok{+}\StringTok{ }\KeywordTok{rnorm}\NormalTok{(n)}
\NormalTok{W <-}\StringTok{ }\NormalTok{A }\OperatorTok{+}\StringTok{ }\KeywordTok{rnorm}\NormalTok{(n)}
\NormalTok{B <-}\StringTok{ }\NormalTok{Z }\OperatorTok{+}\StringTok{ }\KeywordTok{rnorm}\NormalTok{(n)}
\NormalTok{C <-}\StringTok{ }\KeywordTok{rnorm}\NormalTok{(n)}
\NormalTok{X <-}\StringTok{ }\NormalTok{Z }\OperatorTok{+}\StringTok{ }\NormalTok{W }\OperatorTok{+}\StringTok{ }\NormalTok{C }\OperatorTok{+}\StringTok{ }\KeywordTok{rnorm}\NormalTok{(n)}
\NormalTok{Y <-}\StringTok{ }\NormalTok{X }\OperatorTok{+}\StringTok{ }\NormalTok{B }\OperatorTok{+}\StringTok{ }\NormalTok{C }\OperatorTok{+}\StringTok{ }\KeywordTok{rnorm}\NormalTok{(n)}
\NormalTok{D <-}\StringTok{ }\NormalTok{X }\OperatorTok{+}\StringTok{ }\NormalTok{Y }\OperatorTok{+}\StringTok{ }\KeywordTok{rnorm}\NormalTok{(n)}
\end{Highlighting}
\end{Shaded}

We can represent the process that generated the example data as a
graphical model, see figure 1.

\begin{figure}
\centering
\includegraphics{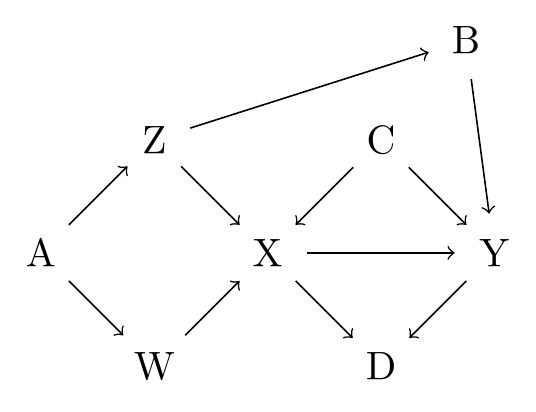}
\caption{Ground truth causal graph of the synthetic data. The variables
\(Z,W,X,Y\) form a \(Y\)-structure after controlling for \(A,B,C\)}
\end{figure}

We will pretend we only will have access to the resulting sample, and
not to the generative equations nor the causal graph. Our goal will be
to verify that the context \(S = \{A,B,C\}\) is a proper control set to
measure the causal effect of the exposure \(X\) and the outcome \(Y\).

Y-tests can help us do this. To verify the proposed control set we aim
to identify an instrument in the data - a cause of the exposure that is
not statistically related to the outcome except through the exposure. In
order to verify that an instrument fulfills this condition, we need an
auxiliary instrument that is independent of the first instrument in the
context \(S\), but becomes dependent after controlling for the exposure.

In our example, we can see in the causal graph that the variables \(Z\)
and \(W\) verify these instrumentality conditions with respect to the
exposure \(X\) and the outcome \(Y\) in the context \(S = \{A,B,C\}\).

In practice, we do not have access to the causal graph. Instead we need
to check these conditions statistically\footnote{There are several ways
  to measure conditional independence - a straightforward one under the
  simplifying assumption of an underlying Gauss-linear system is to
  measure the \(p\)-value of the corresponding coefficient in a least
  squares multivariate regression. Following convention in econometrics
  we will decide that a \(p\)-value under \(0.05\) counts as a measure
  of independence, and we will arbitrarily interpret a \(p\)-value over
  \(0.1\) counts as a measure of dependence; the intermediate range
  corresponds to an uncertain outcome. Note that while the strength of
  the correlation, as measured by e.g.~the standardized regression
  coefficient \(\beta\), is important in determining the importance of a
  correlation, it has no bearing on the statistical \emph{significance}
  of the correlation between variables; hence for our test only the
  \(p\)-value is important. If we cannot make the assumption of
  Gauss-linearity, there are alternative conditional independence tests
  such as those based on conditional mutual information - that case
  however falls outside the scope of the article.}. Concretely, we need
to check that (in context \(S\)) (1) conditioning on \(X\) activates the
correlation between \(Z,W\) and (2) conditioning on \(X\) deactivates
the correlation between \(Z,Y\).

This is not an operational recipe yet. In order to confirm that these
conditions are fulfilled, we need to perform the following subchecks:
(1a) \(Z\) and \(W\) are independent in \(S\), (1b) \(Z\) and \(W\) are
dependent in \(S\cup\{X\}\), (2a) \(Z\) and \(Y\) are conditionally
dependent after controlling for \(S\), and that (2b) \(Z\) and \(Y\) are
conditionally independent after controlling for \(S\cup\{X\}\).

These four checks guarantee that the context \(S\) blocks all non causal
paths between \(X\) and \(Y\). Hence we can (3) measure the causal
effect of \(X\) on \(Y\), by measuring their correlation in context
\(S\).

You can see the results of applying this analysis to our example in
table 1.

\begin{table}[ht]
\centering

\caption{Five steps of the Y test applied to our example.}

\begin{tabular}{@{}llrrr@{}}
\toprule

step & regression & r & sd & p \\\midrule

1a & Z \textasciitilde{} W S & -0.0148 & 0.0734 & 0.8411 \\
1b & Z \textasciitilde{} W \textbar{} X,
S & -0.3708 & 0.0736 & 0.0000 \\
2a & Y \textasciitilde{} Z \textbar{} S & 0.8031 & 0.2316 & 0.0008 \\
2b & Y \textasciitilde{} Z \textbar{} X,S & -0.0707 & 0.1635 & 0.6664 \\
3 & Y \textasciitilde{} X \textbar{} S & 1.0460 & 0.0657 & 0.0000 \\

\bottomrule
\end{tabular}

\end{table}

The two key observations are that (1) a correlation between the
instruments appears after controlling for the exposure and (2) the
correlation between one of the instruments and the outcome disappears
after controlling for the exposure.

These observations constitute a positive result of the Y-test, and
indicate that the context \(S\) does indeed block all non causal paths
from \(X\) to \(Y\).

Speaking inexactly, observation (1) ensures that there is no reverse
path from \(X\) to \(Z\), and the observation (2) ensures that at most
one of \(Z\) and \(Y\) are causes of \(X\).

Furthermore, this proves that \(S\) is a proper control set for
measuring the causal effect of \(X\) on \(Y\). Therefore, the
correlation coefficient on the last regression is an unbiased estimate
of the causal effect of \(X\) on \(Y\).

We provide a formal proof of the soundness of the Y-test in appendix A.

\hypertarget{a-practical-version-of-the-y-test}{%
\subsection{2. A practical version of the
y-test}\label{a-practical-version-of-the-y-test}}

While sound in theory, the y-test is rarely practical because of two
requirements:

\begin{enumerate}
\def\labelenumi{\arabic{enumi}.}
\tightlist
\item
  Identifying a context \(S\) that blocks all non-causal paths between
  \(X\) and \(Y\)
\item
  Identifying two conditionally independent instruments \(Z,W\)
\end{enumerate}

The first requirement forces us to treat y-tests as a test for omitted
variable bias. If we already know there is an unobserved confounder we
cannot block, then the test will invariably fail. Therefore, by
contraposition, if the test returns positive, then we will have
effectively ruled out such confounders.

The second requirement is often hard to meet in practice. Finding a
valid instrument is hard enough - finding two that we can render
conditionally independent is practically impossible.

Therefore, if we hope to apply the y-test in a practical context, we
need to relax this requirement.

The reason why we need the instruments \(Z\) and \(W\) to be
conditionally independent is so we can verify that \(X\) activates the
correlation between \(Z\) and \(W\), and thus rule out the possibility
of a causal path from \(X\) to \(Z\).

Sometimes we can get away with this assumption when we have temporal
information about the variables - in principle, if \(Z\) happened before
\(X\) then we can rule out reverse causation appealing to the temporal
logic of causation. But when our data does not have such a clear
temporal ordering we need an alternative.

Hence we need to find an intermediate test that allows us to distinguish
the cases where \(Z\) is an effect of \(X\) or there is a common cause
for both from the cases where \(Z\) is a cause of \(X\).

In order to find such a test, we study the 8 causal graphs that exhaust
all possible relations between \(Z,W,X\) in a context \(S\), assuming
that 1) the graph between \(Z,X,W\) is complete, 2) \(Z\) and \(W\) are
dependent via a hidden, uncontrollable common cause\footnote{and 3)
  there are no common descendants of the variables we are controlling
  for, explicitly or implicitly}. See figure 2 for a graphical
representation of the selected graphs.

\begin{figure}
\centering
\includegraphics{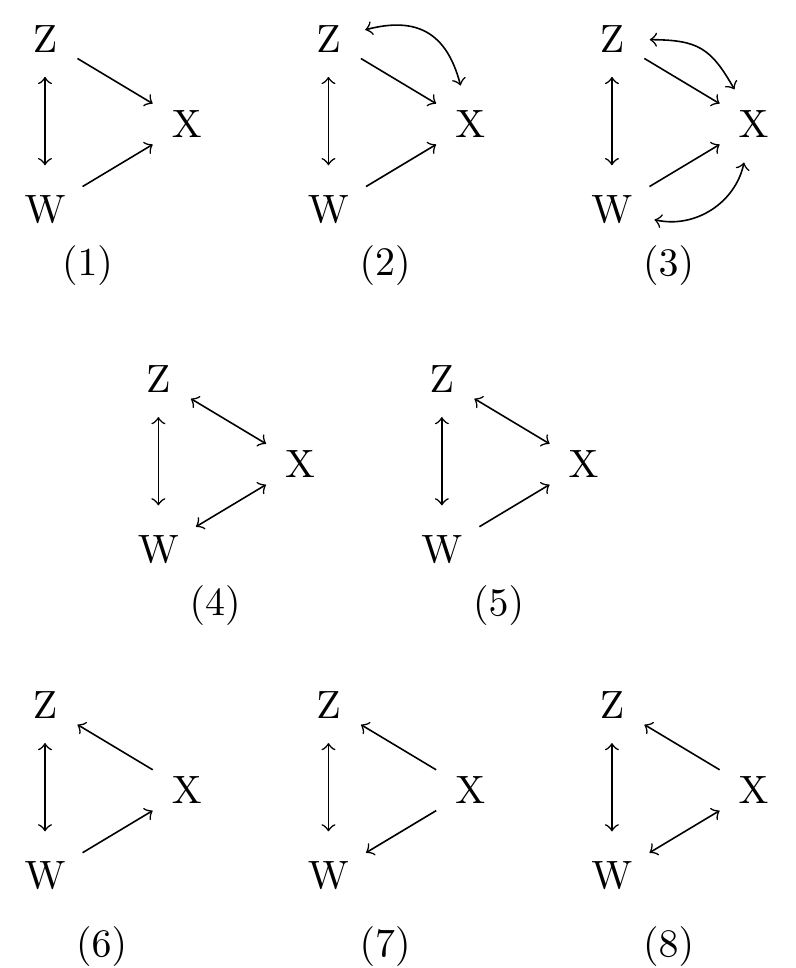}
\caption{Causal graphs we are studying. Single arrows denote a direct
linear causal relation. Double arrows denote a hidden common cause
inducing a spurious correlation between the variables.}
\end{figure}

These 8 graphs include three graphs where there is a path from \(X\) to
either \(Z\) or \(W\) (graphs (6), (7), (8)). This is the situation we
need to rule out for the Y-test to come out positive.

In order to rule out this possible path from \(X\) to \(Z\), we will
study random representatives of each of the 8 graphs we are interested
in. To do this, we define generative functions, which randomly sample
coefficients for the edges in the graph and then sample iid observations
from the resulting causal system.

See below an example of such a generative function for the graph where
both \(Z\) and \(W\) are causes of \(X\) (graph (1)). The definition of
the generative functions corresponding to graphs (2-8) are shown in
appendix B.

\begin{Shaded}
\begin{Highlighting}[]
\NormalTok{sigma <-}\StringTok{ }\DecValTok{5} \CommentTok{# Std of the gaussian noise }

\CommentTok{# Function to sample path coefficients}
\NormalTok{sample_coefficients <-}\StringTok{ }\ControlFlowTok{function}\NormalTok{(n)\{}
\NormalTok{  abs_value <-}\StringTok{ }\KeywordTok{sample}\NormalTok{(}\KeywordTok{c}\NormalTok{(}\OperatorTok{-}\DecValTok{1}\NormalTok{,}\DecValTok{1}\NormalTok{), n,}
                      \DataTypeTok{replace=}\OtherTok{TRUE}\NormalTok{)}
\NormalTok{  sign <-}\StringTok{ }\KeywordTok{sample}\NormalTok{(}\DecValTok{1}\OperatorTok{:}\DecValTok{2}\NormalTok{, n, }\DataTypeTok{replace=}\OtherTok{TRUE}\NormalTok{)}
\NormalTok{  coef <-}\StringTok{ }\NormalTok{abs_value}\OperatorTok{*}\NormalTok{sign}
  \KeywordTok{return}\NormalTok{(coef)}
\NormalTok{\}}

\CommentTok{# Each graph is described by a generative }
\CommentTok{# function that randomly samples its path}
\CommentTok{# coefficients and then returns a sample }
\CommentTok{# of the constructed graph}
\StringTok{"(1)"}\NormalTok{ =}\StringTok{ }\ControlFlowTok{function}\NormalTok{(n)\{}
\NormalTok{  alpha <-}\StringTok{ }\KeywordTok{sample_coefficients}\NormalTok{(}\DecValTok{4}\NormalTok{)}
  
\NormalTok{  A <-}\StringTok{ }\NormalTok{sigma}\OperatorTok{*}\KeywordTok{rnorm}\NormalTok{(n)}
\NormalTok{  Z <-}\StringTok{ }\NormalTok{alpha[}\DecValTok{1}\NormalTok{]}\OperatorTok{*}\NormalTok{A }\OperatorTok{+}\StringTok{ }\KeywordTok{rnorm}\NormalTok{(n)}
\NormalTok{  W <-}\StringTok{ }\NormalTok{alpha[}\DecValTok{2}\NormalTok{]}\OperatorTok{*}\NormalTok{A }\OperatorTok{+}\StringTok{ }\KeywordTok{rnorm}\NormalTok{(n)}
\NormalTok{  X <-}\StringTok{ }\NormalTok{alpha[}\DecValTok{3}\NormalTok{]}\OperatorTok{*}\NormalTok{Z }\OperatorTok{+}\StringTok{ }\NormalTok{alpha[}\DecValTok{4}\NormalTok{]}\OperatorTok{*}\NormalTok{W }\OperatorTok{+}\StringTok{ }\KeywordTok{rnorm}\NormalTok{(n)}
  
\NormalTok{  data <-}\StringTok{ }\KeywordTok{data.frame}\NormalTok{(Z,X,W)}
  
  \KeywordTok{return}\NormalTok{(data)}
\NormalTok{\}}
\end{Highlighting}
\end{Shaded}

Now that we have a way of constructing and sampling random
representatives for each of the possible graphs, we can study different
properties of each of them. The goal will be to find a property that is
true only of graphs where there are no paths from \(X\) to \(Y\).

We focus on (1) whether the correlation between \(Z\) and \(W\) weakens
or becomes stronger after controlling for \(X\) (in terms of its
p-value), and (2) whether the sign of the correlation between \(Z\) and
\(W\) remains the same after controlling for \(X\).

We look at 1000 representatives of each of the 8 graphs, sample 50
independent observations from each representative and check whether
either of the properties (1,2) hold for each sample. The results of the
experiment are summarized in table 2.

\begin{table}[ht]
\centering

\caption{Computational study. For each graph we sample a random
structural representative, and take 50 samples from the representative.
Then we check whether in each sample there is a p value strengthening, a
reversal of the correlation point estimate or either of the two former.}

\begin{tabular}{@{}lrrr@{}}
\toprule

Graph & Strengthening & Reversal & Strengthening or reversal \\\midrule

(1) & 230 & 526 & 756 \\
(2) & 848 & 449 & 962 \\
(3) & 651 & 351 & 797 \\
(4) & 549 & 64 & 594 \\
(5) & 548 & 334 & 779 \\
(6) & 111 & 498 & 518 \\
(7) & 532 & 245 & 561 \\
(8) & 552 & 323 & 632 \\

\bottomrule
\end{tabular}

\end{table}

Let's interpret the results of this exploration.

There are three hypotheses of interest to consider: A)
\texttt{both\ Z\ and\ W\ are\ causes\ of\ X} (graphs G1,G2,G3) B)
\texttt{X\ is\ a\ cause\ of\ either\ Z\ or\ W} (graphs G6,G7,G8) C)
\texttt{\$X\$\ is\ not\ a\ cause\ of\ neither\ \$Z\$\ nor\ \$W\$\ but\ either\ \$X\$\ is\ not\ an\ effect\ of\ \$Z\$\ or\ \$W\$\ is\ not\ a\ cause\ of\ \$X\$}
(graphs G4,G5)

We recall that the hypothesis we wish to rule out is B. If we could rule
it out, then we could apply the second part of the Y-test to conclude
that our control set \(S\) is proper.

We then observe that either there is a decreasing p-value or a sign
reversal (E) or the opposite (\textasciitilde E).

We want to figure out how either result changes our beliefs about the
likelihood of the different hypotheses. In order to do so we need to
follow Bayes rule and look at the likelihood of the evidence given the
different hypothesis: \(P(E|A):P(E|B):P(E|C)\) and
\(P(~E|A):P(~E|B):P(~E|C)\).

Each hypothesis encompasses several possible graph structures, and hence
we have to assign them a relative prior probability beforehand. For
example, we will assign them equal probabilities, except for those
graphs which have a symmetric counterpart after exchanging the role of
\(Z\) and \(W\) (graphs G2, G5, G6 and G8), to which we will assign
twice the prior probability.

Once we have done that, we can resort to our computational study for a
crude estimation of the likelihoods. For example:

\[
P(E|A) = P(E|G_1)P(G_1|A) + P(E|G_2)P(G_2|A) +P(E|G_3)P(G_3|A) \approx 
\frac{756}{1000}` \frac{1}{4} + \frac{962}{1000} \frac{2}{4} + \frac{797}{1000} \frac{1}{4} = 
0.86925
\]

Applying this reasoning to the rest of the likelihoods we finally
conclude that \(P(E|A):P(E|B):P(E|C) \approx 0.189:0.2072:0.198\) and
that \(P(\neg E|A):P(\neg E|B):P(\neg E|C) \approx 0.811:0.7928:0.802\).

Hence a positive observation provides twice as much evidence for
hypothesis A than for hypothesis B, whereas a negative observation
provides 10 times more evidence in favor of hypothesis A relative to
hypothesis B.

This means that we can use the indicator \(E\) as a falsification test
for \(A\). When the indicator is positive it does not provide a strong
signal, so false positives are likely, but a negative is a strong
indication that \(Z\) is a cause of \(X\).

This is to be expected - in the graphs corresponding to hypothesis A,
controlling for the \(X\) opens up another path through which
information can flow, which for typical coefficient values and a
representative sample results in either a stronger correlation or an
outright reversal of the correlation.

Unfortunately the signal is not very useful for distinguishing
hypothesis C from hypothesis B. The likelihood ratio between these given
E is about 1.3 in favor of C, and given \(\neg E\) it is about 2 in
favor of B - not a very strong distinction. But we can live with this
downside: most often, the proposed instruments are both hypothesized to
be causes of the exposure (hypothesis A), and hence hypothesis \(A\) is
the one we are more interested in falsifying.

This concludes our analysis and shapes our new test. Once we have
attempted to falsify that the instrument \(Z\) is a cause of \(X\) in
context \(S\) using this signal, we can proceed to check that the
exposure \(X\) deactivates the correlation between the instrument \(Z\)
and the outcome \(Y\) in context \(S\) to conclude that \(S\) is a
proper control set, and measure the unbiased effect of \(X\) on \(Y\) by
controlling for \(S\).

In the next section we will show how to apply this modified y-test with
real econometric data.

\hypertarget{effect-of-price-on-cigarette-demand}{%
\section{3. Effect of price on cigarette
demand}\label{effect-of-price-on-cigarette-demand}}

As a practical illustration, we consider the problem of modeling the
effect of cigarette price on demand at the US state level. We work with
the same data as is used in Stock and Watson (2011) to illustrate the
application of a generalized two-stage regression model\footnote{A
  reference table of the variables we use in our analysis can be found
  in appendix C}.

This dataset is a good fit for our purposes since (1) it purportedly
includes two instruments, making the Y-test applicable, and (2) there
are strong theoretical reasons to expect that price causally influences
demand and the two instruments shown (two types of taxes) are causes of
the price.

We first load the dataset and replicate the original two stage
regression.

We are working with five variables, aggregated at state level: the
outcome is the number of cigarette packs bought, the exposure is the
price of the cigarettes, the instruments are state-specific general tax
sale and the cigarette tax sale, and the context we control for includes
the income per capita. We work with data from 1995.

In the original analysis an assumption is made that the outcome is
causally downstream from the exposure and that, after controlling for
income, the taxes are proper instruments for the relation we intend to
study. An estimation of the causal effect of the price on demand is made
through a two stage linear regression. We replicate their results in
table X.

\begin{table}[ht]
\centering

\caption{}

\begin{tabular}{@{}lrrrr@{}}
\toprule

term & estimate & std.error & statistic & p.value \\\midrule

(Intercept) & 9.8949555 & 0.9592169 & 10.315660 & 0.0000000 \\
log(rprice) & -1.2774241 & 0.2496100 & -5.117680 & 0.0000062 \\
log(rincome) & 0.2804048 & 0.2538897 & 1.104436 & 0.2752748 \\

\bottomrule
\end{tabular}

\end{table}

Hence, if the causal assumptions made by the authors hold, we conclude
that price and demand are negatively correlated, as predicted by
standard microeconomic theory.

The interesting part that we can add on is that we can attempt to
falsify these causal assumptions using the modified y-test from the last
section.

Again, this will require us to study the contextual relationship between
the instruments before and after controlling for the exposure (steps 1a
and 1b respectively), and the contextual relationship between one of the
instruments and the outcome before and after controlling for the
exposure (steps 2a and 2b respectively). Finally we will report the
contextual correlation between exposure and outcome after controlling
for our candidate control (step 3) The results are summarized in table
@ref(results\_cigs).

\begin{table}[ht]
\centering

\caption{Five steps of the heuristic Y test applied to the data on
cigarette sales.}

\begin{tabular}{@{}llrrr@{}}
\toprule

step & regression & r & sd & p \\\midrule

1a & salestax \textasciitilde{} cigtax \textbar{}
log(rincome) & 0.1721 & 0.0388 & 0.0001 \\
1b & salestax \textasciitilde{} cigtax \textbar{} log(rprice) +
log(rincome) & -0.2407 & 0.0669 & 0.0008 \\
2a & log(packs) \textasciitilde{} cigtax \textbar{}
log(rincome)S & -0.0148 & 0.0034 & 0.0001 \\
2b & log(packs) \textasciitilde{} cigtax \textbar{} log(rprice) +
log(rincome) & 0.0075 & 0.0078 & 0.3415 \\
3 & log(packs) \textasciitilde{} log(rprice) \textbar{}
log(rincome) & -1.4065 & 0.2609 & 0.0000 \\

\bottomrule
\end{tabular}

\end{table}

We observe that the cigarette tax per state, the price of cigarettes per
state and the demand for cigarettes exhibit a chain-like behaviour - the
tax and demand are conditionally dependent (2a), but they become
independent after controlling for the price (2b).

We, however, find that the instruments are positively correlated before
conditioning for the exposure (1a). Hence the traditional y-test would
not be able to detect the causal relationship between price and packs
bought.

But we observe (1) that the instruments exhibit a correlation reversal
after controlling for the exposure - which increases our confidence that
the exposure is an effect of both instruments, as we explain in section
2.

One concern with this analysis is that this reversal could be driven by
outliers. As a simple test to rule out this possibility, we can plot the
data for the two instruments, using color to represent states where the
cigarette prices are similar.

\includegraphics{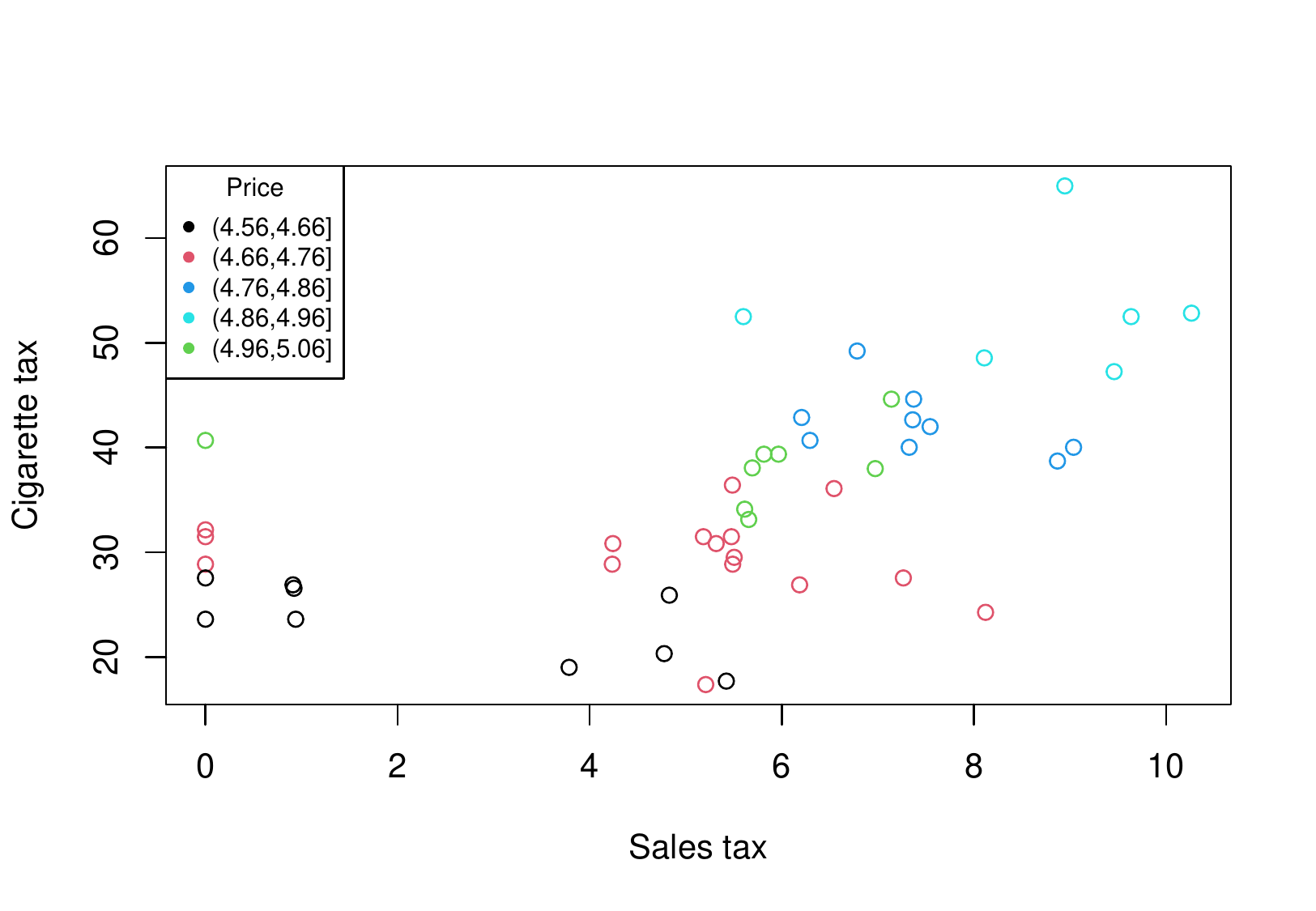}

While there are some outlier states where the sales tax is below 2\% and
another outlier state where the cigarette-specific tax is over 60\%, at
first glance the reversal seems to not be spurious\footnote{If we redo
  the study of the relationship between the instruments after excluding
  these outliers we obtain similar results, with an estimated instrument
  correlation of 0.0971339 {[}0.0259978{]} before controlling for the
  exposure and of -0.1097408 {[}0.0566311{]} after controlling for the
  exposure}.

This data provides moderate evidence that the two instruments, the
exposure and the outcome form a Y-structure, and hence that the packs
bought are causally downstream from the prices and that we are not
missing any important confounders.

\begin{figure}
\centering
\includegraphics{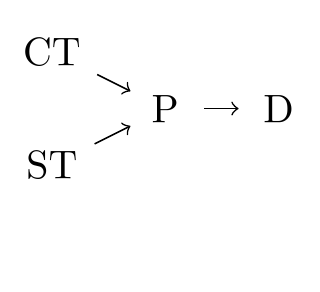}
\caption{Highest likelihood subgraph according to the y-test, relating
the cigarette tax (\(CT\)), the general sales tax \(ST\), the price per
pack of cigarettes (\(P\)) and the demand for cigarettes (\(D\)) after
controlling for income}
\end{figure}

Indeed, we see that the point estimate after controlling for income
falls within the 70\% confidence estimate of the two-stage instrument
regression, which indicates that all important confounders are
controlled for.

\hypertarget{effect-of-firm-loans-on-sales}{%
\section{4. Effect of firm loans on
sales}\label{effect-of-firm-loans-on-sales}}

We saw that our proposed heuristic method for verifying the completeness
of the control set was effective in the context of cigarette prices and
demand. In this section, we apply our test to another use case to
further explore how applicable the proposed method is in practice.

Concretely, we replicate and examine the results of one of the
instrumental analyses performed in `Do firms want to borrow more?'
(Banerjee and Duflo 2014)\footnote{To be exact, we focus on the analysis
  presented in (table 9, column 5) of the original paper. A table with
  the variables used in the analysis for reference can be found in
  appendix C}.

The study is looking at whether firms in India are credit constrained.
It does so by studying the causal effect of the change in log working
capital limit granted, previous to current year, (\texttt{ldgl}) and
their subsequent change of log gross sales (\texttt{dlgsf})\footnote{Other
  proxies for the outcome are considered, but we will focus on this
  particular one for the purpose of illustrating our method}.

Four variables are used as controls: the dummy \texttt{post} is equal to
1 in years 1999 and 2000, zero otherwise. The dummy \texttt{post2} is
equal to 1 in years 2001--2002, zero otherwise. The dummy \texttt{big}
is equal to 1 for firms with a large investment in plant and
machinery\footnote{larger than Rs. 6.5 millions}, zero otherwise. The
dummy \texttt{big2} is equal to 1 for firms with a very large investment
in plant and machinery\footnote{larger than Rs. 10 million}.

The interactions \texttt{post*med}, \texttt{post2*big2} and
\texttt{post*big2} are used as instruments. \texttt{med} is a dummy
indicating that the firm's investment in plant and machinery is
medium\footnote{between Rs. 6.5 million and Rs. 10 million}.

The authors decided to filter some outliers, restricting their analysis
to firms where the response variable \texttt{ldgl} is contained between
-1 and 1.

We replicate Banerjee and Duflo (2014) results in table X. The analysis
is a straightforward two-stage instrumental variable analysis with
multiple instruments, studying the causal relationship between the
variables \texttt{ldgl}, the loan, and \texttt{dlgsf}, gross sales.

\begin{table}[ht]
\centering

\caption{}

\begin{tabular}{@{}lrrrr@{}}
\toprule

term & estimate & std.error & statistic & p.value \\\midrule

(Intercept) & -0.0434255 & 0.1121425 & -0.3872355 & 0.6986960 \\
ldgl & 1.1907750 & 1.0479873 & 1.1362495 & 0.2562292 \\
post & 0.0389093 & 0.0472279 & 0.8238621 & 0.4102900 \\
post2 & -0.0060974 & 0.0557906 & -0.1092908 & 0.9130022 \\
big & 0.0223917 & 0.0534700 & 0.4187709 & 0.6755081 \\
big2 & 0.0354311 & 0.0710304 & 0.4988158 & 0.6180611 \\

\bottomrule
\end{tabular}

\end{table}

Our estimate of the effect of \texttt{ldgl} on \texttt{dlgsf} is
1.190775 (coeftest\_result{[}`ldgl',`Std. Error'{]}). This result does
not exactly match the paper results, but falls within a standard
deviation of their estimate.

Regardless, we want to check the validity of the proposed instruments.
For this we apply the procedure developed in this article, focusing on
two of the instruments, namely the \texttt{post*med} interaction and the
\texttt{post2*big2} interaction\footnote{As there are three instruments
  there are three possible pairings we could be studying. We chose the
  pairing that exhibits the largest change in correlation after
  controlling for the exposure.}. The summary of the results can be
found in table @ref(tab:causality\_test\_dufflo\_results).

\begin{table}[ht]
\centering

\caption{Five steps of the heuristic Y test applied to the data on firm
credit loans.}

\begin{tabular}{@{}llrrr@{}}
\toprule

step & regression & r & sd & p \\\midrule

1a & post\emph{med \textasciitilde{} post2}big2 \textbar{} post + post2
+ big + big2 & -0.0362888 & 0.0122679 & 0.0031727 \\
1b & post\emph{med \textasciitilde{} post2}big2 \textbar{} ldgl + post +
post2 + big + big2 & -0.0345640 & 0.0124107 & 0.0054588 \\
2a & dlgsf \textasciitilde{} post2*big2 \textbar{} post + post2 + big +
big2 & -0.1336647 & 0.1168405 & 0.2530057 \\
2b & dlgsf \textasciitilde{} post2*big2 \textbar{} ldgl + post + post2 +
big + big2 & -0.1118817 & 0.1186749 & 0.3461204 \\
3 & dlgsf \textasciitilde{} ldgl \textbar{} post + post2 + big +
big2 & 0.2004837 & 0.0729202 & 0.0061205 \\

\bottomrule
\end{tabular}

\end{table}

We find that the two of the instruments we focus on (namely
\texttt{post*med} and \texttt{post2*big2}) have a strong correlation,
r=-0.0362888(0.0122679). p=0.0031727. Furthermore, they remain
correlated at a lower level of significance, and exhibit no reversal,
after controlling for the exposure, at r=-0.034564(0.0124107).
p=0.0054588. This suggests that the instruments are not both causal
ancestors of the exposure, since we would not expect that to happen in
typical v-structures.

Furthermore, the correlation between the outcome \texttt{dlgsf} and the
instrument \texttt{ldgl} is barely affected by conditioning on the
exposure. This suggests that either the main relationship between the
instrument \texttt{ldgl} and the outcome \texttt{dlgsf} is not mediated
through the exposure or there is an unobserved confounder.

The above observations seem to weaken the case for (Banerjee and Duflo
2014) analysis. Particularly, the absence of a significant instrument
correlation after controlling for the exposure seems to invalidate the
implicit causal assumptions of the IV analysis conducted.

However we can build a toy model following the author's assumptions that
replicates the phenomena observed in the data quite faithfully. This is
evidence that our heuristic is misleading in this case.

\begin{Shaded}
\begin{Highlighting}[]
\CommentTok{# Example of a data generation process}
\CommentTok{# where there heuristic Y-test fails}
\NormalTok{n <-}\StringTok{ }\DecValTok{100}
\NormalTok{z1 <-}\StringTok{ }\KeywordTok{rnorm}\NormalTok{(n) }\OperatorTok{<}\StringTok{ }\FloatTok{0.} 
\NormalTok{z2 <-}\StringTok{ }\KeywordTok{rnorm}\NormalTok{(n) }\OperatorTok{<}\StringTok{ }\FloatTok{1.}
\NormalTok{z3 <-}\StringTok{ }\KeywordTok{rnorm}\NormalTok{(n) }\OperatorTok{<}\StringTok{ }\FloatTok{-1.}
\NormalTok{i1 <-}\StringTok{ }\NormalTok{z1}\OperatorTok{*}\NormalTok{z2}
\NormalTok{i2 <-}\StringTok{ }\NormalTok{z2}\OperatorTok{*}\NormalTok{z3}
\NormalTok{e <-}\StringTok{ }\NormalTok{i1 }\OperatorTok{+}\StringTok{ }\NormalTok{i2 }\OperatorTok{+}\StringTok{ }\NormalTok{z1 }\OperatorTok{+}\StringTok{ }\NormalTok{z2 }\OperatorTok{+}\StringTok{ }\NormalTok{z3 }\OperatorTok{+}\StringTok{ }\KeywordTok{rnorm}\NormalTok{(n)}
\NormalTok{o <-}\StringTok{ }\NormalTok{e }\OperatorTok{+}\StringTok{ }\NormalTok{z1 }\OperatorTok{+}\StringTok{ }\NormalTok{z2 }\OperatorTok{+}\StringTok{ }\NormalTok{z3 }\OperatorTok{+}\StringTok{ }\KeywordTok{rnorm}\NormalTok{(n)}
\end{Highlighting}
\end{Shaded}

\begin{verbatim}
## [1] "i1 ~  i2 | s1 + s2 + s3 = 0.05 [0.27], p=0.8445588759"
\end{verbatim}

\begin{verbatim}
## [1] "i1 ~  i2 | e + s1 + s2 + s3 = 0.06 [0.27], p=0.8265246505"
\end{verbatim}

The result in this simulation is that, as with the real dataset, the
correlation of the instrument does not vary much after controlling for
the exposure ; executing the piece of code above a few times will
convince the reader that this result is not atypical.

Hence we conclude that our heuristic test is not guaranteed to be
informative outside of the strict Gauss linear systems with
approximately uniform distribution of coefficients where we developed
our heuristic in section 2.

Characterizing scenarios where the heuristic is unreliable remains an
open question; for example in this dataset it may be because of the use
of binary interactions as instruments or because the contextual
independences present in the data.

\hypertarget{conclusion-and-open-questions}{%
\section{5. Conclusion and open
questions}\label{conclusion-and-open-questions}}

In this article, we have explained how to use the Y-test to identify
causal relationships using two conditionally independent instruments.

Under the reasoning that this test is too restrictive in practice, we
have done a computational study of causal graphs that helped us identify
a weaker requirement than conditionally independent instruments. We show
that if the strength of the correlation between the instruments
diminishes after controlling for the exposure but there is no reversal,
that is a strong indication that at least one of the instruments is not
a cause of the exposure.

Finally, we have shown how to apply this modified y-test in the context
of a price-demand elasticity problem.

The practicality of this test is yet in question. For example, in
section 4 we have shown an example of a dataset where our proposed
heuristic is misleading. Understanding better where it is appropriate to
rely on this statistic is an important next step.

The high rate of false positives is also concerning, but it is to be
expected in such low-data scenarios, so we could argue that, in that
respect, the proposed test is not different from other econometric
tools. Of course, there are tools which have better asymptotic
properties that we cannot guarantee for our test - whether the test will
be misleading depends primarily on the underlying causal structure, not
the amount of data collected.

While the test we propose is still quite restrictive on its conditions
of application (as it requires the identification of two presumed
instruments), it successfully relaxes the stringent condition of
instrument independence required by the Y-test. We hope this heuristic
will be a useful concept for the econometrician's toolbox - primarily as
a test to falsify a proposed control set.

\hypertarget{future-work}{%
\subsection{5.1 Future work}\label{future-work}}

In principle, the Y-test could be extended to deal with possibly
non-linear, non-gaussian processes by using an appropriate conditional
independence test.

Another question we haven't considered is whether this reasoning can be
extended to contexts with possibly cyclical causal dependencies, and
context where there is selection bias.

We also might be interested in considering the possibility of multiple
Y-tests supporting contradictory results. Naively, we could take them as
independent evidence that cancels out - but this is unsatisfactory, as
we would expect that different tests provide different strength of
evidence.

Lastly, while the Y-test is sound, it is not complete - there are causal
relationships that can be determined from conditional independence tests
that cannot be detected applying the Y-test. Hence we could use more
general principles of causal discovery such as the ones exposed in Zhang
(2008) or Claassen and Heskes (2011) to build more general tests.

\hypertarget{references}{%
\section{References}\label{references}}

\hypertarget{refs}{}
\leavevmode\hypertarget{ref-banerjee_firms_2014}{}%
Banerjee, Abhijit V., and Esther Duflo. 2014. ``Do Firms Want to Borrow
More? Testing Credit Constraints Using a Directed Lending Program.''
\emph{The Review of Economic Studies} 81 (2): 572--607.
\url{https://doi.org/10.1093/restud/rdt046}.

\leavevmode\hypertarget{ref-claassen_structure_2011}{}%
Claassen, Tom, and Tom Heskes. 2011. ``A Structure Independent Algorithm
for Causal Discovery.'' In \emph{In ESANN'11}, 309--14.

\leavevmode\hypertarget{ref-glymour_review_2019}{}%
Glymour, Clark, Kun Zhang, and Peter Spirtes. 2019. ``Review of Causal
Discovery Methods Based on Graphical Models.'' \emph{Frontiers in
Genetics} 10. \url{https://doi.org/10.3389/fgene.2019.00524}.

\leavevmode\hypertarget{ref-mani_theoretical_2012}{}%
Mani, Subramani, Peter L. Spirtes, and Gregory F. Cooper. 2012. ``A
Theoretical Study of Y Structures for Causal Discovery.''
\emph{arXiv:1206.6853 {[}Cs, Stat{]}}, June.
\url{http://arxiv.org/abs/1206.6853}.

\leavevmode\hypertarget{ref-messerli_chocolate_2012}{}%
Messerli, Franz H. 2012. ``Chocolate Consumption, Cognitive Function,
and Nobel Laureates.'' \emph{New England Journal of Medicine} 367 (16):
1562--4. \url{https://doi.org/10.1056/NEJMon1211064}.

\leavevmode\hypertarget{ref-reiersol_confluence_1945}{}%
Reiersöl, Olav. 1945. ``Confluence Analysis by Means of Instrumental
Sets of Variables.''
\url{/paper/Confluence-analysis-by-means-of-instrumental-sets-Reiers\%C3\%B6l/e24fab07b33f9521febf11ccdbea0d5bdb38f927}.

\leavevmode\hypertarget{ref-sargan_estimation_1958}{}%
Sargan, J. D. 1958. ``The Estimation of Economic Relationships Using
Instrumental Variables.'' \emph{Econometrica} 26 (3): 393--415.
\url{https://doi.org/10.2307/1907619}.

\leavevmode\hypertarget{ref-stock_introduction_2011}{}%
Stock, James H., and Mark Watson. 2011. \emph{Introduction to
Econometrics: International Edition}. 3rd edition. Boston, Mass.:
Pearson Education.

\leavevmode\hypertarget{ref-zhang_completeness_2008}{}%
Zhang, Jiji. 2008. ``On the Completeness of Orientation Rules for Causal
Discovery in the Presence of Latent Confounders and Selection Bias.''
\emph{Artificial Intelligence} 172 (16): 1873--96.
\url{https://doi.org/10.1016/j.artint.2008.08.001}.

\hypertarget{a.-proof-of-the-soundness-of-the-y-test}{%
\section{A. Proof of the soundness of the
Y-test}\label{a.-proof-of-the-soundness-of-the-y-test}}

Before we can state the theorem we need some preliminary definitions.

Definition 1 - Causal structural model

A \textbf{structural causal model} consists of a set of variables
\(X = \{X_1, ..., X_n\}\), noise terms \(U_1, ..., U_n\) and a set of
functional constraints \(X_i = f_i(pa(X_i), U_i)\), where
\(pa(X_i) \subset X\) is a subset of the variables called the
\emph{parents} or \emph{direct causes} of \(X_i\).

Associated to a causal structural model we have a \textbf{causal graph},
which is a directed graph \(G = <X,E>\) whose nodes are the variables
\(X\) and there is an edge from \(X_i\) to \(X_j\) iff
\(X_i\in pa(X_j)\).

We say that a structural causal model is \textbf{acyclic} if the
parenthood relation induces an acyclic causal graph.

Definition 2 - d-separation and faithfulness

We say that a trail \(A_1 - A_2 - ... - A_n\) in a directed acyclic
graph \(G = <X,E>\) is blocked in a context \(S\) iff either (1) it
contains a chain \(A_{i-1}\rightarrow A_i\rightarrow A_{i+1}\) such that
\(A_i\in S\), (2) it contains a fork
\(A_{i-1}\leftarrow A_i\rightarrow A_{i+1}\) such that \(A_i\in S\) or
(3) it contains a collider \(A_{i-1}\rightarrow A_i\leftarrow A_{i+1}\)
such that neither \(A_i\) nor any of its descendants are in \(S\).

We say that two variables \(A,B\) in a directed acyclic graph are
\textbf{d-separated} by a context \(S\subset X\) iff every trail between
\(A,B\) is blocked by \(S\).

We say that two variables \(X_i, X_j\) in a structural causal model are
conditionally independent given a context \(S\subset X\) iff \(X_i\) and
\(X_j\) are conditionally independent for any possible assignment for
the context \(S=s\). We write this as \(X_i \perp X_j | S\).

We say that an acyclic structural causal model is \textbf{faithful} when
there is a one-to-one correspondence between d-separation in the causal
graph and conditional independence. That is, when \(X_i \perp X_j | S\)
iff \(X_i\) and \(X_j\) are d-separated by \(S\) in its associated
causal graph.

Theorem 1 - Soundness of the y-test

Let's assume that we have four variables \(A,B,Z,W\) and a context \(S\)
in an acyclical and faithful causal structural model such that: 1.
\(Z\perp W | S\) and \(Z\perp W | S\cup \{A\}\) 2.
\(Z \not \perp B | S\) and \(Z \perp B | \cup \{A\}\) 3.
\(A \not \perp B | S\)

Then we can conclude that \(A\) is a cause of \(B\), ie there exists a
fully directed path \(A\to B\) in the causal diagram associated to the
system.

Furthermore, \(S\) is a proper controlling set for \(A\to B\) - it
blocks all and only the non causal paths from \(A\) to \(B\).

\emph{Proof sketch} -

Our goal is to show that there exists a directed path \(A\to B\) in the
causal graph.

Let \(Z - W\) be a trail between \(Z\) and \(W\) that becomes active in
context \(S\cup \{A\}\). Because of d-separation properties, in this
trail \(A\) is either a collider or the children of a collider. In
either case, it is true that there exists an active trail \(Z\to_SA\)
such that the last edge points towards \(A\).

Now, let \(A-_SB\) be any active trail between \(A,B\) in context \(S\)
(we know that such a trail must exist since we assumed that
\(A \not \perp B\) in context \(S\)).

Consider the concatenated trail \(Z\to_S A -_S B\), which we know is
active in \(S\). Because of our assumption that \(Z \not \perp B | S\)
and \(Z \perp B | \cup \{A\}\), any active trail between \(Z, A, B\) in
\(S\) becomes blocked in context \(S\cup \{A\}\). Hence, we know that
\(A\) is not a collider in this particular trail. Since we already know
that there is an incident edge \(Z\to_S A\) in the trail, we must
conclude that the first edge in the trail \(A -_S B\) points away from
\(A\).

Hence, either \(A-_S B\) is a directed path from \(A\to B\) or there
exists a collider in the path that is activated by \(S\).

But suppose that there existed a collider in the trail \(A-_S B\). Pick
whichever collider is closer to \(A\). For the trail to be active, the
collider must be active, so either the collider itself is in \(S\) or a
direct descendant of the collider is in \(S\). Either way, since the
closest collider to \(A\) is a direct descendant of \(A\), then we have
that a direct descendant of \(A\) is in \(S\).

But that cannot be, since \(A\) was either a deactivated collider or the
direct descendant of a deactivated collider in the trail \(Z-_S W\).
Hence we conclude that every active trail \(A-_S B\) is a fully directed
path from \(A\) to \(B\). In particular, there exists some fully
directed path \(A\to B\), hence \(A\) is a cause of \(B\).

Additionally, we can show that no fully directed trail from \(A\to B\)
is blocked by \(S\). Indeed, if it was otherwise, when there would be a
descendant \(C\) of \(A\) such that \(C\in S\). But then \(C\) would be
a common descendant of \(Z\) and \(W\) in \(S\), and we would have that
\(Z \not\perp W | S\), against our premise. Hence \(S\) is a proper
control set for \(A\to B\) - it blocks all and only the non causal
trails from \(A\) to \(B\).

\(\square\)

\hypertarget{b.-definition-of-the-generative-functions-to-sample-graphs}{%
\section{B. Definition of the generative functions to sample
graphs}\label{b.-definition-of-the-generative-functions-to-sample-graphs}}

The code below is used to generate random graphs from which to sample in
section 2.

\begin{Shaded}
\begin{Highlighting}[]
\CommentTok{# Function to randomly sample path coefficients}
\NormalTok{sample_coefficients <-}\StringTok{ }\ControlFlowTok{function}\NormalTok{(n)\{}
\NormalTok{  abs_value <-}\StringTok{ }\KeywordTok{sample}\NormalTok{(}\KeywordTok{c}\NormalTok{(}\OperatorTok{-}\DecValTok{1}\NormalTok{,}\DecValTok{1}\NormalTok{), n, }\DataTypeTok{replace=}\OtherTok{TRUE}\NormalTok{)}
\NormalTok{  sign <-}\StringTok{ }\KeywordTok{sample}\NormalTok{(}\DecValTok{1}\OperatorTok{:}\DecValTok{2}\NormalTok{, n, }\DataTypeTok{replace=}\OtherTok{TRUE}\NormalTok{)}
\NormalTok{  coef <-}\StringTok{ }\NormalTok{abs_value}\OperatorTok{*}\NormalTok{sign}
  \KeywordTok{return}\NormalTok{(coef)}
\NormalTok{\}}

\CommentTok{# List of representative causal graphs}
\CommentTok{# Each graph is described by a generative }
\CommentTok{# function that randomly samples its path}
\CommentTok{# coefficients and then returns a sample }
\CommentTok{# of the constructed graph}
\NormalTok{graphs <-}\StringTok{ }\KeywordTok{list}\NormalTok{(}
    \StringTok{"(1)"}\NormalTok{ =}\StringTok{ }\ControlFlowTok{function}\NormalTok{(n)\{}
\NormalTok{    alpha <-}\StringTok{ }\KeywordTok{sample_coefficients}\NormalTok{(}\DecValTok{4}\NormalTok{)}
    
\NormalTok{    A <-}\StringTok{ }\NormalTok{sigma}\OperatorTok{*}\KeywordTok{rnorm}\NormalTok{(n)}
\NormalTok{    Z <-}\StringTok{ }\NormalTok{alpha[}\DecValTok{1}\NormalTok{]}\OperatorTok{*}\NormalTok{A }\OperatorTok{+}\StringTok{ }\KeywordTok{rnorm}\NormalTok{(n)}
\NormalTok{    W <-}\StringTok{ }\NormalTok{alpha[}\DecValTok{2}\NormalTok{]}\OperatorTok{*}\NormalTok{A }\OperatorTok{+}\StringTok{ }\KeywordTok{rnorm}\NormalTok{(n)}
\NormalTok{    X <-}\StringTok{ }\NormalTok{alpha[}\DecValTok{3}\NormalTok{]}\OperatorTok{*}\NormalTok{Z }\OperatorTok{+}\StringTok{ }\NormalTok{alpha[}\DecValTok{4}\NormalTok{]}\OperatorTok{*}\NormalTok{W }\OperatorTok{+}\StringTok{ }\KeywordTok{rnorm}\NormalTok{(n)}
    
\NormalTok{    data <-}\StringTok{ }\KeywordTok{data.frame}\NormalTok{(Z,X,W)}
    
    \KeywordTok{return}\NormalTok{(data)}
\NormalTok{  \},}
  
  \StringTok{"(2)"}\NormalTok{ =}\StringTok{ }\ControlFlowTok{function}\NormalTok{(n)\{}
\NormalTok{    alpha <-}\StringTok{ }\KeywordTok{sample_coefficients}\NormalTok{(}\DecValTok{6}\NormalTok{)}
    
\NormalTok{    A <-}\StringTok{ }\KeywordTok{rnorm}\NormalTok{(n)}
\NormalTok{    B <-}\StringTok{ }\KeywordTok{rnorm}\NormalTok{(n)}
\NormalTok{    Z <-}\StringTok{ }\NormalTok{alpha[}\DecValTok{1}\NormalTok{]}\OperatorTok{*}\NormalTok{A }\OperatorTok{+}\StringTok{ }\NormalTok{alpha[}\DecValTok{2}\NormalTok{]}\OperatorTok{*}\NormalTok{B }\OperatorTok{+}\StringTok{ }\KeywordTok{rnorm}\NormalTok{(n)}
\NormalTok{    W <-}\StringTok{ }\NormalTok{alpha[}\DecValTok{3}\NormalTok{]}\OperatorTok{*}\NormalTok{A }\OperatorTok{+}\StringTok{ }\KeywordTok{rnorm}\NormalTok{(n)}
\NormalTok{    X <-}\StringTok{ }\NormalTok{alpha[}\DecValTok{4}\NormalTok{]}\OperatorTok{*}\NormalTok{B }\OperatorTok{+}\StringTok{ }\NormalTok{alpha[}\DecValTok{5}\NormalTok{]}\OperatorTok{*}\NormalTok{W }\OperatorTok{+}\StringTok{ }\NormalTok{alpha[}\DecValTok{6}\NormalTok{]}\OperatorTok{*}\NormalTok{Z }
         \OperatorTok{+}\StringTok{ }\KeywordTok{rnorm}\NormalTok{(n)}
    
\NormalTok{    data <-}\StringTok{ }\KeywordTok{data.frame}\NormalTok{(Z,X,W)}
    
    \KeywordTok{return}\NormalTok{(data)}
\NormalTok{  \},}

  \StringTok{"(3)"}\NormalTok{ =}\StringTok{ }\ControlFlowTok{function}\NormalTok{(n)\{}
\NormalTok{    alpha <-}\StringTok{ }\KeywordTok{sample_coefficients}\NormalTok{(}\DecValTok{8}\NormalTok{)}
    
\NormalTok{    A <-}\StringTok{ }\KeywordTok{rnorm}\NormalTok{(n)}
\NormalTok{    B <-}\StringTok{ }\KeywordTok{rnorm}\NormalTok{(n)}
\NormalTok{    C <-}\StringTok{ }\KeywordTok{rnorm}\NormalTok{(n)}
\NormalTok{    Z <-}\StringTok{ }\NormalTok{alpha[}\DecValTok{1}\NormalTok{]}\OperatorTok{*}\NormalTok{A }\OperatorTok{+}\StringTok{ }\NormalTok{alpha[}\DecValTok{2}\NormalTok{]}\OperatorTok{*}\NormalTok{B }\OperatorTok{+}\StringTok{ }\KeywordTok{rnorm}\NormalTok{(n)}
\NormalTok{    W <-}\StringTok{ }\NormalTok{alpha[}\DecValTok{3}\NormalTok{]}\OperatorTok{*}\NormalTok{A }\OperatorTok{+}\StringTok{ }\NormalTok{alpha[}\DecValTok{4}\NormalTok{]}\OperatorTok{*}\NormalTok{C }\OperatorTok{+}\StringTok{ }\KeywordTok{rnorm}\NormalTok{(n)}
\NormalTok{    X <-}\StringTok{ }\NormalTok{alpha[}\DecValTok{5}\NormalTok{]}\OperatorTok{*}\NormalTok{B }\OperatorTok{+}\StringTok{ }\NormalTok{alpha[}\DecValTok{6}\NormalTok{]}\OperatorTok{*}\NormalTok{C }\OperatorTok{+}\StringTok{ }\NormalTok{alpha[}\DecValTok{7}\NormalTok{]}\OperatorTok{*}\NormalTok{Z }
         \OperatorTok{+}\StringTok{ }\NormalTok{alpha[}\DecValTok{8}\NormalTok{]}\OperatorTok{*}\NormalTok{W }\OperatorTok{+}\StringTok{ }\KeywordTok{rnorm}\NormalTok{(n)}
    
\NormalTok{    data <-}\StringTok{ }\KeywordTok{data.frame}\NormalTok{(Z,X,W)}
    
    \KeywordTok{return}\NormalTok{(data)}
\NormalTok{  \},}
  
  \StringTok{"(4)"}\NormalTok{ =}\StringTok{ }\ControlFlowTok{function}\NormalTok{(n)\{}
\NormalTok{    alpha <-}\StringTok{ }\KeywordTok{sample_coefficients}\NormalTok{(}\DecValTok{6}\NormalTok{)}
    
\NormalTok{    A <-}\StringTok{ }\KeywordTok{rnorm}\NormalTok{(n)}
\NormalTok{    B <-}\StringTok{ }\KeywordTok{rnorm}\NormalTok{(n)}
\NormalTok{    C <-}\StringTok{ }\KeywordTok{rnorm}\NormalTok{(n)}
\NormalTok{    Z <-}\StringTok{ }\NormalTok{alpha[}\DecValTok{1}\NormalTok{]}\OperatorTok{*}\NormalTok{A }\OperatorTok{+}\StringTok{ }\NormalTok{alpha[}\DecValTok{2}\NormalTok{]}\OperatorTok{*}\NormalTok{B }\OperatorTok{+}\StringTok{ }\KeywordTok{rnorm}\NormalTok{(n)}
\NormalTok{    W <-}\StringTok{ }\NormalTok{alpha[}\DecValTok{3}\NormalTok{]}\OperatorTok{*}\NormalTok{A }\OperatorTok{+}\StringTok{ }\NormalTok{alpha[}\DecValTok{4}\NormalTok{]}\OperatorTok{*}\NormalTok{C }\OperatorTok{+}\StringTok{ }\KeywordTok{rnorm}\NormalTok{(n)}
\NormalTok{    X <-}\StringTok{ }\NormalTok{alpha[}\DecValTok{5}\NormalTok{]}\OperatorTok{*}\NormalTok{B }\OperatorTok{+}\StringTok{ }\NormalTok{alpha[}\DecValTok{6}\NormalTok{]}\OperatorTok{*}\NormalTok{C }\OperatorTok{+}\StringTok{ }\KeywordTok{rnorm}\NormalTok{(n)}
    
\NormalTok{    data <-}\StringTok{ }\KeywordTok{data.frame}\NormalTok{(Z,X,W)}
    
    \KeywordTok{return}\NormalTok{(data)}
\NormalTok{  \},}
  
  \StringTok{"(5)"}\NormalTok{ =}\StringTok{ }\ControlFlowTok{function}\NormalTok{(n)\{}
\NormalTok{    alpha <-}\StringTok{ }\KeywordTok{sample_coefficients}\NormalTok{(}\DecValTok{5}\NormalTok{)}
    
\NormalTok{    A <-}\StringTok{ }\KeywordTok{rnorm}\NormalTok{(n)}
\NormalTok{    B <-}\StringTok{ }\KeywordTok{rnorm}\NormalTok{(n)}
\NormalTok{    Z <-}\StringTok{ }\NormalTok{alpha[}\DecValTok{1}\NormalTok{]}\OperatorTok{*}\NormalTok{A }\OperatorTok{+}\StringTok{ }\NormalTok{alpha[}\DecValTok{2}\NormalTok{]}\OperatorTok{*}\NormalTok{B }\OperatorTok{+}\StringTok{ }\KeywordTok{rnorm}\NormalTok{(n)}
\NormalTok{    W <-}\StringTok{ }\NormalTok{alpha[}\DecValTok{3}\NormalTok{]}\OperatorTok{*}\NormalTok{A }\OperatorTok{+}\StringTok{ }\KeywordTok{rnorm}\NormalTok{(n)}
\NormalTok{    X <-}\StringTok{ }\NormalTok{alpha[}\DecValTok{4}\NormalTok{]}\OperatorTok{*}\NormalTok{B }\OperatorTok{+}\StringTok{ }\NormalTok{alpha[}\DecValTok{5}\NormalTok{]}\OperatorTok{*}\NormalTok{W }\OperatorTok{+}\StringTok{ }\KeywordTok{rnorm}\NormalTok{(n)}
    
\NormalTok{    data <-}\StringTok{ }\KeywordTok{data.frame}\NormalTok{(Z,X,W)}
    
    \KeywordTok{return}\NormalTok{(data)}
\NormalTok{  \},}
  
  \StringTok{"(6)"}\NormalTok{ =}\StringTok{ }\ControlFlowTok{function}\NormalTok{(n)\{}
\NormalTok{    alpha <-}\StringTok{ }\KeywordTok{sample_coefficients}\NormalTok{(}\DecValTok{4}\NormalTok{)}
    
\NormalTok{    A <-}\StringTok{ }\KeywordTok{rnorm}\NormalTok{(n)}
\NormalTok{    W <-}\StringTok{ }\NormalTok{alpha[}\DecValTok{1}\NormalTok{]}\OperatorTok{*}\NormalTok{A }\OperatorTok{+}\StringTok{ }\KeywordTok{rnorm}\NormalTok{(n)}
\NormalTok{    X <-}\StringTok{ }\NormalTok{alpha[}\DecValTok{2}\NormalTok{]}\OperatorTok{*}\NormalTok{W }\OperatorTok{+}\StringTok{ }\KeywordTok{rnorm}\NormalTok{(n)}
\NormalTok{    Z <-}\StringTok{ }\NormalTok{alpha[}\DecValTok{3}\NormalTok{]}\OperatorTok{*}\NormalTok{A }\OperatorTok{+}\StringTok{ }\NormalTok{alpha[}\DecValTok{4}\NormalTok{]}\OperatorTok{*}\NormalTok{X }\OperatorTok{+}\StringTok{ }\KeywordTok{rnorm}\NormalTok{(n)}
    
\NormalTok{    data <-}\StringTok{ }\KeywordTok{data.frame}\NormalTok{(Z,X,W)}
    
    \KeywordTok{return}\NormalTok{(data)}
\NormalTok{  \},}

  \StringTok{"(7)"}\NormalTok{ =}\StringTok{ }\ControlFlowTok{function}\NormalTok{(n)\{}
\NormalTok{    alpha <-}\StringTok{ }\KeywordTok{sample_coefficients}\NormalTok{(}\DecValTok{4}\NormalTok{)}
    
\NormalTok{    A <-}\StringTok{ }\KeywordTok{rnorm}\NormalTok{(n)}
\NormalTok{    X <-}\StringTok{ }\KeywordTok{rnorm}\NormalTok{(n)}
\NormalTok{    W <-}\StringTok{ }\NormalTok{alpha[}\DecValTok{1}\NormalTok{]}\OperatorTok{*}\NormalTok{A }\OperatorTok{+}\StringTok{ }\NormalTok{alpha[}\DecValTok{2}\NormalTok{]}\OperatorTok{*}\NormalTok{X }\OperatorTok{+}\StringTok{ }\KeywordTok{rnorm}\NormalTok{(n)}
\NormalTok{    Z <-}\StringTok{ }\NormalTok{alpha[}\DecValTok{3}\NormalTok{]}\OperatorTok{*}\NormalTok{A }\OperatorTok{+}\StringTok{ }\NormalTok{alpha[}\DecValTok{4}\NormalTok{]}\OperatorTok{*}\NormalTok{X }\OperatorTok{+}\StringTok{ }\KeywordTok{rnorm}\NormalTok{(n)}
    
\NormalTok{    data <-}\StringTok{ }\KeywordTok{data.frame}\NormalTok{(Z,X,W)}
    
    \KeywordTok{return}\NormalTok{(data)}
\NormalTok{  \},}

  \StringTok{"(8)"}\NormalTok{ =}\StringTok{ }\ControlFlowTok{function}\NormalTok{(n)\{}
\NormalTok{    alpha <-}\StringTok{ }\KeywordTok{sample_coefficients}\NormalTok{(}\DecValTok{5}\NormalTok{)}
    
\NormalTok{    A <-}\StringTok{ }\KeywordTok{rnorm}\NormalTok{(n)}
\NormalTok{    B <-}\StringTok{ }\KeywordTok{rnorm}\NormalTok{(n)}
\NormalTok{    W <-}\StringTok{ }\NormalTok{alpha[}\DecValTok{1}\NormalTok{]}\OperatorTok{*}\NormalTok{A }\OperatorTok{+}\StringTok{ }\NormalTok{alpha[}\DecValTok{2}\NormalTok{]}\OperatorTok{*}\NormalTok{B }\OperatorTok{+}\StringTok{ }\KeywordTok{rnorm}\NormalTok{(n)}
\NormalTok{    X <-}\StringTok{ }\NormalTok{alpha[}\DecValTok{3}\NormalTok{]}\OperatorTok{*}\NormalTok{B }\OperatorTok{+}\StringTok{ }\KeywordTok{rnorm}\NormalTok{(n)}
\NormalTok{    Z <-}\StringTok{ }\NormalTok{alpha[}\DecValTok{4}\NormalTok{]}\OperatorTok{*}\NormalTok{A }\OperatorTok{+}\StringTok{ }\NormalTok{alpha[}\DecValTok{5}\NormalTok{]}\OperatorTok{*}\NormalTok{X }\OperatorTok{+}\StringTok{ }\KeywordTok{rnorm}\NormalTok{(n)}
    
\NormalTok{    data <-}\StringTok{ }\KeywordTok{data.frame}\NormalTok{(Z,X,W)}
    
    \KeywordTok{return}\NormalTok{(data)}
\NormalTok{  \}}
\NormalTok{)}
\end{Highlighting}
\end{Shaded}

\hypertarget{c.-list-of-variables-used-in-the-analysis}{%
\section{C. List of variables used in the
analysis}\label{c.-list-of-variables-used-in-the-analysis}}

Variables used for the analysis in section 3:

\begin{table}[ht]
\centering

\caption{}

\begin{tabular}{@{}ll@{}}
\toprule

Variable & Meaning \\\midrule

salestax & Percentage amount of general tax over goods in a state \\
cigtax & Percentage amount of tax over cigarettes in a state \\
rprice & Price of a cigarette pack, adjusted for inflation \\
rincome & Average income per capita, adjusted for inflation \\

\bottomrule
\end{tabular}

\end{table}

Variables used for the analisis in section 4:

\begin{table}[ht]
\centering

\caption{}

\begin{tabular}{@{}ll@{}}
\toprule

Variable & Meaning \\\midrule

dlgsf & Change in log working capital limit granted, previous to current
year \\
ldgl & Change of log gross sales, previous to current year \\
post & Indicator of whether the year is 1999-2000 \\
post2 & Indicator of whether the year is 2001-2002 \\
big & Indicator of whether investment in plant and machinery is larger
than Rs. 6.5 million \\
big2 & Indicator of whether investment in plant and machinery is larger
than Rs. 10 million \\
med & Indicator of whether investment in plant and machinery is between
than Rs. 6.5 million and Rs. 10 million \\

\bottomrule
\end{tabular}

\end{table}

\end{document}